\documentstyle[preprint,revrex]{aps}
\begin{document}
\draft
\preprint{}
\begin{title}
 { QCD At Finite Temperature - A Variational Approach}
 \end{title}
 \author{A. Mishra}
 \begin{instit}
{Physics Department, Utkal University, Bhubaneswar-751004, India}
\end{instit}
\author { H. Mishra, S.P. Misra, S.N. Nayak}
\begin{instit}
{Institute of Physics, Bhubaneswar-751005, India.}
\end{instit}
\begin{abstract}
We develop here a nonperturbative framework to study
quantum chromodynamics (QCD) at finite temperatures
using the thermofield dynamics (TFD) method of Umezawa. The methodology
considered here
is self-consistent and variational. There is a dynamical generation
of a magnetic gluon mass. This eliminates the infrared problems
associated with perturbative QCD calculations at finite
temperatures. We obtain here the thermodynamical quantities
like free energy density, pressure and entropy density.
We also calculate the temperature dependence of
SVZ parameter ${\alpha_{s} \over \pi}<{G^a}_{\mu \nu }
{G^{a \mu \nu }}>$. The condensate
vanishes at the critical temperature in accordance with recent hot sum
rule calculations. The present method gives an insight to the
vacuum structure in QCD at zero temperature as well as at
finite temperatures in a co-ordinated manner.
\end{abstract}
\pacs{}
\narrowtext
\section{INTRODUCTION}

Vacuum structure in quantum chromodynamics (QCD) is nonperturbative
and highly nontrivial \cite{sav77}.
We have attempted the same for zero temperature earlier \cite{am91}
through a nonperturbative variational method, where
the scale parameter of QCD vacuum was related to
that of Shifman, Vainshtein and
Zakharov (SVZ) in the context of sum rules \cite{svz}. We shall now
extend the same to QCD at finite temperature with the thermofield dynamics
method of Umezawa \cite{tfd}

It is well known that QCD at finite temperatures remains nonperturbative
due to serious infrared divergences \cite{linde80}. This makes the perturbative
expansion ineffective \cite{anishet84} unless an appropriate
 partial resummationis done \cite{pis90}. The problem has also
  been dealt with a semiclassical
expansion around a gas of chromomagnetic monopoles \cite{polya77}.
Although the QCD electric propagators are screened by Debye
mass at one loop level, the magnetic channel remains long range
at any finite order of perturbation theory \cite{nadkarni86}.
As we have noted in Ref. 2, a mass gap in the magnetic sector
naturally arises through the non trivial vacuum structure
of gluon condensates. We shall use thermofield dynamics to
generalise the vacuum structure to finite temperature.

We organise the paper as follows. In section 2 we consider
the algebra for
the ground state structure of QCD at finite temperature.
For this purpose, we briefly recapitulate the same for zero
temperature \cite{am91},  and then consider temperature dependent
field theory with thermofield dynamics by minimising
free energy. Here we study the temperature dependence
of the gluon condensates describing the structure of
vacuum \cite{tfd} as well as the same for other parameters.
The scale which was determined through SVZ parameter \cite{am91},
is here used to write the temperature dependence of the
physical quantities.In section 3 we discuss the
free energy extremisation and obtain the results for different
thermodynamic quantities. The critical temperature here
appears to be about 275 MeV, with the $selfconsistent$
gluon mass being around 300-400 MeV.
In section 4, we summarize and discuss the present
methodology.

The method is tested for solvable models \cite{hm88} and, as
stated, for QCD at zero \hfill temperature \cite{am91} with results similar to
that of earlier methods \cite{cornwal82}.
It  has also been applied for ground state structure and other problems
in nuclear physics \cite{np}.
 It is however conceptually different and
has more predictive power regarding offshell boson quanta
with testable conclusions elsewhere \cite{higgs}.

\section { QCD at finite temperature}

We shall consider QCD at a finite temperature using thermofield
methods \cite{tfd}  For this purpose, we
start with the QCD Lagrangian given as
\begin{equation}
{\cal L} =-{1\over 2}G^{a\mu\nu}(\partial_{\mu}{W^{a}}_{\nu}
-\partial_{\nu}{W^{a}}_{\mu}+gf^{abc}{W^{b}}_{\mu}{W^{c}_{\nu})}
+{1\over 4}{G^{a}}_{\mu\nu}{G^{a\mu\nu}},
\end{equation}
\noindent where ${W^{a}}_{\mu}$ are the SU(3) color gauge fields.
We shall quantise in Coulomb gauge and write the
electric field  ${G^{a}}_{0i}$ in terms of
the transverse and longitudinal parts as
\begin{equation}
{G^a}_{0i}=
{^TG^a}_{0i}+{\partial_{i}{f}^{a}},
\end{equation}
\noindent where the form of ${f}^{a}$ is to be determined.
In the Coulomb gauge the
subsidiary condition and the equal time algebra for the gauge
fields are given as \cite{schwinger}
\begin{equation}
{\partial_{i}}{W^a}_i=0
\end{equation}
\noindent and
\begin{equation}
\left [{W^{a}}_{i}({\vec x},t),^{T}{G^{b}}_{0j}
({\vec y},t)\right ]=i{\delta}^{ab}({\delta}_{ij}
-{{\partial_{i}\partial_{j}}\over {\partial^2}})\delta
({\vec x}-{\vec y}).
\end{equation}
\noindent We take the field
expansions for ${W^{a}}_{i}$ and $^{T}{G^{a}}_{0i}$
at time t=0 as \cite{shut85}
\begin{mathletters}
\begin{equation}
{W^{a}}_{i}(\vec x)={(2\pi)^{-3/2}}\int
{d\vec k\over \sqrt{2\omega(\vec k)}}({a^a}_{i}(\vec k) +
{{a^a}_{i}(-\vec k)}^{\dagger})\exp({i\vec k.\vec x})
\end{equation}
 \noindent and
 \begin{equation}
 {^{T}G^{a}}_{0i}(\vec x)={(2\pi)^{-3/2}} i \int
{d\vec k}{\sqrt{\omega(\vec k)\over 2}}(-{a^a}_{i}(\vec k) +
{{a^a}_{i}(-\vec k)}^{\dagger})\exp({i\vec k.\vec x}).
\end{equation}
\end{mathletters}
\noindent From equation (4) these give the commutation
relations for ${a^a}_{i}$ and ${{a^b}_{j}}^\dagger$ as
\begin{equation}
\left[ {a^a}_{i}(\vec k),{{a^b}_{j}(\vec k^{'})}^\dagger\right]=
\delta^{ab}\Delta_{ij}(\vec k)\delta({\vec k}-{\vec k^{'}}),
\end{equation}
\noindent where, $\omega$(k) is arbitrary \cite{shut85}, and,
 \begin{equation}
 \Delta_{ij}(\vec k)={\delta_{ij}}-{k_{i}k_{j}\over k^2}.
\end{equation}
\noindent In Coulomb gauge, the expression for the Hamiltonian
density, ${\cal T}^{00}$ from equation (1) is given as \cite{schwinger}
\begin{equation}
{\cal T}^{00}=:{1\over 2}{^{T}{G^a}_{0i}}{^{T}{G^a}_{0i}}+
{1\over 2}{W^a}_{i}(-\vec \bigtriangledown^2){W^a}_{i}+
{{g^2}\over 4}f^{abc}f^{aef}{W^b}_{i}{W^c}_{j}{W^e}_{i}{W^f}_{j}+
{1\over 2}(\partial_{i}f^{a})(\partial_{i}f^{a}):,
\end{equation}
\noindent where : : denotes the normal ordering with respect to
the perturbative vacuum, say $\mid vac>$, defined through
${a^a}_{i}(\vec k)\mid vac>=0$. In equation (8) we have not retained
the cubic terms in gauge fields {\it since with gluon pair condensates
these will not contribute}.
This implies omission of some possible contributions to the vacuum
structure. In fact, the cubic term will contribute to vacuum structure
provided
we accept gluon condensates consisting of {\em three} gluons. We
have omitted it for the sake of simplicity.
The term ${1 \over 2}(\partial _if^a)
(\partial _if^a)$ automatically includes interactions for both
time-like and longitudinal gluons, here through the auxiliary field
description, and ${\cal T}^{00}$ is calculated after this elimination
through equations (9) and (10) given later.
We note that \cite{am91}
\begin{equation}
f^a=-{W^a}_0-g \; f^{abc}\;{ (\vec \bigtriangledown ^2)}^{-1}
({W^b}_i \; \partial _i {W^c}_0).
\end{equation}
\noindent Hence, eliminating $f^a$, the equation for ${W^a}_0$ becomes
\begin{equation}
\vec \bigtriangledown ^2{W^a}_0 + 2g \; f^{abc}{W^b}_i \; \partial _i
{W^c}_0 + g^2 \; f^{abc}f^{cde} \; {W^b}_i \partial _i
((\vec \bigtriangledown ^2)^{-1}({W^d}_j \partial _j{W^e}_0))
={J^a}_0,
\end{equation}
\noindent where
\begin{equation}
{J^a}_{0}=gf^{abc}{W^b}_{i}^{T}{G^c}_{0i}.
\end{equation}
It is not possible to solve the above equation
for ${W^a}_0$.
We shall therefore proceed with a mean field type of approximation
 for the ground state. In thermofield method \cite{tfd}
 the ground state is written as $\mid vac';\beta>$ with $\beta=1/T$ and
 it is a state in an extended Hilbert space including thermal modes. We
 shall therefore replace the left hand side of equation (10)
 with the expectation values for $\mid vac';\beta>$ for all
the fields other than ${W^a}_0$. Then, the above equation gets replaced by
\begin{eqnarray}
\vec \bigtriangledown ^2{W^a}_0 (\vec x )
&&+ g^2 \; f^{abc}f^{cde} \;<vac',\beta\mid  {W^b}_i(\vec x ) \partial _i
(\vec \bigtriangledown ^2)^{-1}({W^d}_j(\vec x ) \mid vac',\beta>
\partial _j{W^e}_0(\vec x ))\nonumber\\ && ={J^a}_0(\vec x ).
\end{eqnarray}

We note that at zero temperature, $\mid vac';\beta=\infty>=\mid vac'>$
was the nonpertubative ground state as discussed in Ref.2. It will be
instructive to recapitulate the expressions of Ref.2 that will get
modified at finite temperature.
We define  $\mid vac'>$ through a unitary
transformation, in a similar manner to
Gross-Neveu model considered earlier \cite{hm88}, given as
\begin{equation}
\mid vac ^{'}>
=U\mid vac>,
\end{equation}
 \noindent where
 \begin{equation}
 U=\exp({B^\dagger}-B),
\end{equation}
\noindent  The unitary operator U for the temperature dependent case
will be discussed later.
In Ref.2, it was shown that at zero temperature, we may have
\begin{equation}
{B^\dagger}={1\over 2}
\int {f(\vec k){{{a^a}_{i}(\vec k)}^\dagger}
{{{a^a}_{i}(-\vec k)}^{\dagger}}d\vec k},
\end{equation}
\noindent
where $f(\vec k)$ describes gluon condensates. It was also
shown \cite{am91} that a nonperturbative vacuum with $f(\vec k)\not =0$
was favoured above a critical coupling.
As noted earlier, we are not including, e.g., three gluon
condensates for the vacuum structure in our ansatz.
Temperature dependence will arise with $U$ above depending
on temperature and the Hilbert space being doubled.
For a moment we continue to take zero temperature.
With the above transformation, the operators, say
${b^a}_{i}(\vec k)$, which annihilate $\mid vac^{'}>$ are
given as
\begin{mathletters}
\begin{equation}
{b^a}_{i}(\vec k)=U{a^a}_{i}(\vec k){U}^{-1}.
\end{equation}
\noindent We explicitly evaluate from the above equation
the operators ${b^a}_{i}(\vec k)$
corresponding to the state, $\mid vac^{'}>$ as related to the
operators corresponding to the state, $\mid vac>$ through
the Bogoliubov transformation given as
\begin{equation}
\left(
\begin{array}{c}
 {b^a}_{i}(\vec k) \\ {{b^a}_{i}(-\vec k)}^\dagger
 \end{array}
 \right)
=\left(
\begin{array}{cc}
 \cosh f(\vec k) & -\sinh f(\vec k) \\
-\sinh f(\vec k) & \cosh f(\vec k)
\end{array}
\right)
\left(
\begin{array}{c}
{a^a}_{i}(\vec k)  \\
{a^a}_{i}(-\vec k)^\dagger,
\end{array}
\right)
\end{equation}
\end{mathletters}
\noindent where the function
 $f(\vec k)$  is even in $\vec k$ and has been assumed to be real.
\noindent Using equations (6) and (16b), we obtain the same
commutation relation for the operators ${b^a}_{i}$
and ${{b^b}_{j}}^\dagger$ given as
\begin{eqnarray}
\left[ {b^a}_{i}(\vec k),
{{b^b}_{j}(\vec k^{'})}^\dagger \right]
& = & \delta^{ab}\Delta_{ij}(\vec k)\delta({\vec k}-{\vec k^{'}}),
\nonumber
\end{eqnarray}
 which merely reflects that the Bogoliubov
transformation (16b) is a canonical transformation.
It is useful to define
\begin{equation}
<vac' \mid :{W^a}_{i}(\vec x){W^b}_{j}(\vec y):\mid  vac'>=
{\delta }^{ab}
\times (2  \pi )^{-3}\int d\vec k e^{i\vec k.(\vec x-
\vec y)}\; {F_{+}(\vec  k)\over \omega (k)}\;
\Delta _{ij}(\vec k),\end{equation}
\begin{equation}{<vac' \mid}: {^{T} G^{a}_{0i}} (\vec x),
{^{T} G^{b}_{0j}} (\vec y):{\mid vac'>} = \delta ^{ab} (2 \pi )^{-3}
\int d{\vec k}e^{i{\vec k}.{(\vec x-\vec y)}}
{\Delta _{ij}(\vec k)\over\omega (k)} F_{-}( k).\end{equation}
In the above $F_{+}(k)$ and $F_{-}(k)$ are given as
\begin{mathletters}
\begin{equation}F_{+}(\vec k)=\biggl (
{\sinh 2f(k)\over 2}
+ {\sinh}^{2}f(k)
\biggr ),\end{equation}
\begin{equation}F_{-}(\vec k)=\biggl ({\sinh}^2 f(k)
- {\sinh 2f(k)\over 2}
\biggr ).\end{equation}
\end{mathletters}
Using the above it was earlier seen that the solution for
$\tilde W^a \! _0 (\vec k)$ is given by
\begin{equation}
(\vec k ^2+ \phi (\vec k ))\tilde W^a \! _0(\vec k )=
-\tilde J^a \! _0(\vec k ),
\end{equation}
 \noindent where
$\tilde W^a \! _0 (\vec k)$ and $\tilde J^a \! _0 (\vec k)$
are Fourier transforms of $W^a \! _0 (\vec x)$
and $J^a \! _0 (\vec x)$; and, with $\;{ g^2/ 4 \pi }=\alpha _s$,
\begin{equation}\phi (\vec k )=
{3 \alpha _s \over 2 \pi } \int {F_+(k')\over {\omega (k')}}dk'
\Big[ (k^2+k'^2)-
{ (k^2-k'^2)^2
\over 2 kk'}\times ln \Big |{k+k'
\over k-k'} \Big | \Big].
\end{equation}
 \noindent We note that $f(\vec k)$ occuring in the Bogoliubov
 transformations describing the possible vacuum structure is to
 be determined through minimisation of energy density at zero
 temperature. We may achieve this in simple cases \cite{hm88}, but here it
 is impossible to do so. Hence we adopt the alternative procedure of
 taking a reasonably simple ansatz for $f(\vec k)$ and extremise it
 over parameters in $f(\vec k)$. Since the gluon correlation function
 $f(\vec k)$ should go to zero for large k (condensation being a long
 distance effect) and since in the above equation hyperbolic
functions enter, we choose the simple form, with $k=\mid\vec k\mid$,
 \begin{mathletters}
 \begin{equation}
 \sinh f(\vec k)=Ae^{-Bk^{2}/2},\end{equation}
\noindent where the parameter $A$ is determined through
energy minimisation and the dimensional parameter $B$
gets determined by the SVZ parameter \cite{svz}.
We note that we then have from equation (16b)
\begin{equation}
<vac'\mid {a^a}_{i}(\vec k)^{\dagger}{a^a}_{i}(\vec k')\mid
vac'>=8\times 2 \times A^{2} e^{-Bk ^2}\delta (\vec k -\vec k')
\end{equation}
\end{mathletters}
\noindent so that we are really taking a gaussian distribution function
for the $perturbative$ gluons as a natural ansatz. Here we are aiming at
an understanding of
vacuum structure in the context of low energy phenomenology,
including the properties of the same at finite temperature
as done later.
It is obvious that such a solution will be applicable only for
problems involving low energy physics, which is our present objective.

 Using the form (22a) for $f(\vec k)$, the energy density,
$\epsilon_{0}$ can be written in terms of the dimensionless
quantities $x={\sqrt {B}}k$ and $\mu={\sqrt B}m$ as \cite{am91}
\begin{eqnarray}
\epsilon_{0}
& = & {1\over {B^2}}(I_{1}+I_{2}+{I_{3}}^{2}+I_{4})\nonumber \\
& & \equiv {1\over {B^2}}F(A),
\end{eqnarray}
\noindent where
\begin{mathletters}
\begin{equation}
I_{1}
={4\over {\pi^2}}\int {\omega(x)x^{2}dx
(A^{2}e^{-x^2}-Ae^{-{{x^2}/2}}{(1+A^{2}e^{-x^2})}^{1\over 2})},
\end{equation}
\begin{equation}
I_{2}=
{4\over {\pi^2}}\int {{x^{4}}\over \omega(x)}dx
(A^{2}e^{-x^2}+Ae^{-{{x^2}/2}}{(1+A^{2}e^{-x^2})}^{1\over 2}),
\end{equation}
\begin{equation}
I_{3}=
{{2g}\over {\pi^2}}\int {x^{2}\over \omega(x)}dx
(A^{2}e^{-x^2}+Ae^{-{{x^2}/2}}{(1+A^{2}e^{-x^2})}
^{1\over 2}),\end{equation}
\noindent and
\begin{equation}
I_{4} =4\times {(2 \pi )^{-6}}\int d{\vec x}
{G(\vec x)\over {x^2+\phi (x)}}
\end{equation}
\end{mathletters}
\noindent with
\begin{eqnarray}
 G(\vec x) & = & {3g^{2}}
\int d \vec x'
\biggl (A^{2}e^{-{x'}^2}+Ae^{-{{x'}^2}/2}
{(1+A^{2}e^{-{x'}^2})}^{1\over 2}\biggr )\nonumber \\ & \times
& \biggl (A^{2}e^{-{(\vec x+\vec x')}^2}-
Ae^{-{{(\vec x +\vec x')}^2}/2}
{(1+A^{2}e^{-{(\vec  x+\vec x')}^2})}^{1\over 2}\biggr )\nonumber \\
& \times &
{\omega (\mid \vec x +\vec x'\mid)\over {\omega ( x')}}
\biggl (1 +{{{({x'}^2 +\vec x .\vec x')}^2} \over
{\vec {x'}^2 (\vec x + \vec x')^2}}\biggr ) ,
\end{eqnarray}
\noindent  and
\begin{eqnarray}
 \phi  (x) & = & {3g^2\over  {8 \pi  ^2}}\int dx'
 \biggl (  x^2+{x'}^2-{(x^2-{x'}^2)^2
\over{2xx'}}\log \Big | {{x+x'}\over{x-x'}}\Big | \biggr )
\nonumber \\ & \times & \biggl (A^{2}e^{-{x'}^2}+Ae^{-{{{x'}^2}/2}}
{(1+A^{2}e^{-{x'}^2})}^{1\over 2}\biggr ).
\end{eqnarray}
\noindent In the above, $\omega(x)={(x^2+\mu^2)}^{1\over 2}$,
with $\mu$ as the effective gluon mass in ${1\over {\sqrt B}}$
units. Mass here is merely an effective parameter for the
field expansions in equations (5) and is {\em not}
the energy of the asymptotic particle which do not exist
in view of confinement. Gluon mass has been used here in
that sense, and can not be identified with the masses for
 magnetic and electric gluons. This mass parameter here
is an artifact of the approximation scheme.
We identify the gluon mass $\mu$ from
the sum of the single contractions of the
quartic interaction term of ${\cal T}^{00}$ in equation (8),
the negative of which gives a mass term in the effective Lagrangian.
We thus have the $self consistency$ $ requirement$ that
\begin{equation}
\mu^2={{2g^2}\over {\pi^2}}
\int {{x^2}dx\over \omega(x)}
(A^{2}e^{-x^2}+Ae^{-{{x^2}/2}}{(1+A^{2}e^{-x^2})}
^{1\over 2}).
\end{equation}
\noindent $\mu$ is determined through an iterative procedure
for any particular value of A so that equation (27)
 is satisfied with the input $\mu$ on the right hand side
 giving rise to the same output $\mu$ on the left hand side.
We found that there exists
a critical value $g_c$ of $g$ with ${{g_c}^{2}}/4\pi\simeq 0.39$,
such that for $g > g_{c}$, $A_{min}\not= 0$ and
$\epsilon_{0}$ becomes negative demonstrating instability of
perturbative vacuum.
Two remarks here may be in order. Firstly we note that we have
taken here $g$ as the coupling constant in the Lagrangian.
In semiperturbative QCD, through renormalisation group equation
the coupling constant of the Lagrangian together with the scale
for renormalisation generate the mass scale $\Lambda$ of QCD,
which describes the running coupling constant. At the present
level of only discussing the vacuum energy such a mathematical
structure has not been arrived at. For this purpose, we have to consider
higher order Greens functions with nonperturbative vacuum structure
so that we are in a position to consider the renormalised coupling
constant at different scales and see how it runs. The problem is
nontrivial and has not been addressed here. The second remark we
wish to make is the fact that $g_c > 0$ may be an artifact of the
approximation scheme. We encountered such a situation for Gross-Neveu
model in Ref.10.

 The value of B, which is a scale parameter,
is now determined by relating it with the SVZ parameter.
In fact, the vacuum structure
 of QCD is given as \cite{svz}
\begin{equation}
{{g^2}\over {4{\pi}^2}}<:{G^a}_{\mu\nu}
{G^{a\mu\nu}:>_{vac^{'}}=0.012GeV^{4}},
\end{equation}
\noindent where $\mid vac^{'}>$ is the physical
vacuum. The left hand side of the above equation can be
explicitly evaluated as
\begin{equation}
{{g^2}\over {4{\pi}^2}}<:{G^a}_{\mu\nu}
{G^{a\mu\nu}:>_{vac^{'}}=
{1\over {B^2}}{{g^2}\over {\pi^2}}
{(-I_{1}(A)+I_{2}(A)+I_{3}(A)^{2}-I_{4}(A))\Big |
_{A=A_{min}}}},
\end{equation}
\noindent where $A_{min}$ is the value of A
corresponding to minimised energy density
and using equation (28), the scale parameter, $B$
gets determined as a function of coupling constant. We may note that
here the contributions from $I_2$ and $I_3$ correspond to the
magnetic sector, and, that of $I_1$ and $I_4$, to the
electric sector. In fact, here the magnetic sector contributes
$0.00585 GeV^4$ and the electric sector contributes $0.00615 GeV^4$ for
coupling constant $\alpha_s=0.8$.

We shall now study the effect of temperature on the
above expressions.
For this purpose, we shall use the methodology
of thermofield dynamics \cite{tfd},
with the $``$thermal vacuum" $\mid vac^{'};\beta>$
being defined so as to yield correct distributions for bosons and
fermions. This methodology consists in the doubling of the
Hilbert space, introducing fresh $``$tilde" space with operators
${{\tilde b}^a}_{i}(\vec k)$ and
${{{\tilde b}^a}_{i}(\vec k)}^{\dagger}$ for this space \cite{tfd}.
In particular, thermal vacuum is defined as
\begin{equation}
\mid vac^{'};\beta>
=U(\beta)\mid vac^{'}>,\end{equation}
 \noindent where \begin{equation}U(\beta)
=\exp{({B(\beta)}^{\dagger}-B(\beta))},
\end{equation}
 with
 \begin{equation}
 {B(\beta)}^{\dagger}=
\int \theta(\vec k,\beta){{b^a}_{i}(\vec k)}^{\dagger}
{{{\tilde b}^a}_{i}(-\vec k)}^{\dagger}d\vec k.
\end{equation}
\noindent The above operators
${{\tilde b}^a}_{i}$ and
${{\tilde b}^{b^\dagger}}_{j}$ satisfy the same commutation
algebra as
${{b}^a}_{i}$ and
${{{b}^a}_{i}}^{\dagger}$.
Clearly in the ansatz for the thermal $``$vacuum" we have the
unknown function $\theta (k,\beta )$. This function we shall
calculate through a self consistency principle through a temperature
dependance of the gluon mass. We take as earlier \cite{tfd,higgs}

\begin{equation}
{\sinh}^{2}\theta({\vec k},\beta)=
{1\over {\exp{{(\beta\omega({\vec k},\beta)})}-1}},
\end{equation}
where, for ${\omega({\vec k},\beta)}$ we take the ansatz
\begin{equation}
{\omega(k,{\beta})}={(k^{2}+{m_{eff}
(\beta)}^{2})}^{1/ 2},
\end{equation}
with $m_{eff}(\beta)$ being a temperature dependant mass.
This mass parameter will be determined in a $ self consistent$ manner.
The reason for choosing
$\theta (k,\beta )$ as  in equation (33) is that $\sinh ^2(\theta )$
corresponds to the number operator $({b_{i}^{a}}^{\dagger}b_{i}^{a})$
expectation value in the thermal vacuum and we are taking
the $form$ of free fields parametrised as above.

With the ansatz (30) for the thermal vacuum the operators corresponding to
$\mid vac^{'};\beta>$ are given as \begin{equation}
\left(
 \begin{array}{c}
 {b^a}_{i}(\vec
k,\beta) \\ {{\tilde b}^a}_{i}(-\vec k,\beta)^{\dagger}
\end{array}\right)
=
U(\beta)\left(
\begin{array}{c}
{b^a}_{i}(\vec
k) \\ {{\tilde b}^a}_{i}(-\vec k)^{\dagger}
\end{array}\right)U(\beta)^{-1}.
\end{equation}
\noindent On simplification of the above we see that
the operators corresponding to the \break\hfil
non-perturbative vacuum,
$\mid vac^{'}>$  at $\beta=\infty$, are related to the operators
corresponding to the thermal vacuum, $\mid vac^{'};\beta>$
through the transformation
 \begin{equation}
\left(
 \begin{array}{c}
 {b^a}_{i}(\vec k) \cr {{\tilde b}^a}_{i}(-\vec k)^{\dagger}
\end{array}\right) =
\left(
\begin{array}{cc} \cosh \theta(\vec k,\beta)\; & \;
\sinh \theta(\vec k,\beta) \\
\sinh \theta(\vec k,\beta)\; & \;\cosh \theta(\vec k,\beta)
\end{array}\right)
\left(
\begin{array}{c}
{b^a}_{i}(\vec k,\beta)\\
{{{\tilde b}^{a}}_{i}(-\vec k,\beta)}^\dagger.
\end{array}
\right)
\end{equation}
\noindent Thus,using equations (13) and (36), we see that the
operators corresponding to the perturbative vacuum, ${\mid vac>}$,
are related to the operators corresponding to the thermal vacuum,
$\mid vac^{'};\beta>$ through the transformation
\begin{eqnarray}
\left(
\begin{array}{c}
{a^a}_{i}(\vec k) \\ {{a^a}_{i}(-\vec k)}^{\dagger}
\\ {{\tilde b}^{a}}_{i}(\vec k)
\\ {{{\tilde b}^{a}}_{i}(-\vec k)}^{\dagger}
\end{array} \right) & = & \left(
\begin{array} {cccc}
\cosh f\cosh \theta\; &
\;\sinh f\cosh \theta\; & \;
\sinh f\sinh \theta\; & \;
\cosh f\sinh \theta \\
\sinh f\cosh \theta\; & \;
\cosh f\cosh \theta\; & \;
\cosh f\sinh \theta\; & \;
\sinh f\sinh \theta \\
0 & 0 & \cosh \theta & \sinh \theta \\
0 & 0 & \sinh \theta  &\cosh \theta
\end{array}
\right)\nonumber\\ & \times &
\left(
\begin{array}{c}
{b^a}_{i}(\vec k,\beta) \\
{b^a}_{i}(-\vec k,\beta)^{\dagger} \\
{{\tilde b}^{a}}_{i}(\vec k,\beta) \\
{{\tilde b}^{a}}_{i}(-\vec k,\beta)^{\dagger}.
\end{array}
\right)
\end{eqnarray}
 Our next job is to evaluate the expectation value of
${\cal T}^{00}$ with respect to $\mid vac^{'};\beta>$.
For this purpose, we note that equations (17) and (18)
now get changed to
\begin{equation}
<vac';\beta \mid :{W^a}_{i}(\vec x){W^b}_{j}(\vec y):
\mid  vac';\beta >=
{\delta }^{ab}
\times (2  \pi )^{-3}\int d\vec k e^{i\vec k.(\vec x-
\vec y)}\; {F_{+}(\vec  k,\beta )\over \omega (k,\beta )}\;
\Delta _{ij}(\vec k),
\end{equation}
\begin{equation}
{<vac';\beta \mid}: {^{T} G^{a}_{0i}} (\vec x),
{^{T} G^{b}_{0j}} (\vec y):{\mid vac';\beta >}
= \delta ^{ab} (2 \pi )^{-3}
\int d{\vec k}e^{i{\vec k}.{(\vec x-\vec y)}}
{\Delta _{ij}(\vec k)\over\omega (k,\beta )}
F_{-}( k,\beta ).
\end{equation}
In the above the temperature dependant
 $F_{+}(k,\beta )$ and $F_{-}(k,\beta )$ are given as
 \begin{mathletters}
\begin{eqnarray}
F_{+}(\vec k,\beta ) & = & \biggl ( {\sinh}^{2}f(k)
{\Bigl(1+2 \sinh ^2 \theta (k,\beta )\Bigr)}+\sinh ^2 \theta (k,\beta )
\nonumber\\ & + &
{\sinh 2f(k)\over 2}
\Bigl(1+2 \sinh ^2 \theta (k,\beta )\Bigr)\biggr ),
\end{eqnarray}
\begin{eqnarray}
F_{-}(\vec k,\beta) & = & \biggl ({\sinh}^2 f(k)
\Bigl (1+2 \sinh ^2 \theta (k,\beta )\Bigr )+\sinh ^2 \theta (k,\beta )
\nonumber \\ &
- &  {\sinh 2f(k)\over 2}
{\Bigl (1+2 \sinh ^2 \theta (k,\beta )\Bigr)}
\biggr ).
\end{eqnarray}
\end{mathletters}
 We may verify that when $\beta \rightarrow \infty$, $\theta(k,\beta)
\rightarrow 0$ and equations (40) reduce to equations (19), as expected.
Using equations (8), (38) and (39), we then obtain
the expectation value of ${\cal T}^{00}$ with respect to
$\mid vac^{'};\beta>$ as
\begin{eqnarray}
\epsilon_{0}(\beta) &
\equiv & <vac^{'};\beta \mid
:{\cal T}^{00}:\mid vac^{'};\beta> \nonumber \\
& = & C_{1}(\beta)+C_{2}(\beta)+{C_{3}(\beta)}^{2}+C_{4}(\beta),
\end{eqnarray}
\noindent where
\begin{mathletters}
\begin{eqnarray}
C_{1}(\beta) & = & <:{1\over 2}
{^T}{G^a}_{0i}{^T}{G^a}_{0i}:>_{vac^{'};\beta}\nonumber \\
& = & {4\over {\pi^2}}\int \omega(k)k^{2} F_{-}(k,\beta )\;dk,
\end{eqnarray}
\begin{eqnarray}
C_{2}(\beta ) & = & <:{1\over 2}
{W^a}_{i}{(-\vec \bigtriangledown^2)}{W^a}_{i}:>_{vac';\beta }\nonumber
\\ & = & {4\over {\pi^2}}\int {{k^{4}}\over \omega(k)}\;F_{+}(k,\beta )\;dk
\end{eqnarray}
\begin{eqnarray}
  {C_{3}(\beta )}^{2} & = & <:{1\over 4}g^{2}f^{abc}f^{aef}
{W^b}_{i}{W^c}_{j}{W^e}_{i}{W^f}_{j}:>_{vac';\beta }\nonumber \\
& = & \left({{2g}\over {\pi^2}}\int {{k^{2}}\over
{\omega(k,\beta )}}\;F_{+}(k,\beta )\;dk\right)^2 ,
\end{eqnarray}
\noindent and
\begin{eqnarray}
 C_{4}(\beta ) & = & <:{1\over 2}
(\partial_{i}f^{a})(\partial_{i}f^{a}):>_{vac{'};\beta },\nonumber\\
 & = & 4\times (2 \pi )^{-6}\int  d \vec  k {G(\vec k,\beta )\over
 {k^2+\phi (k,\beta )}},
 \end{eqnarray}
 \end{mathletters}
  \noindent  where
\begin{eqnarray}
G(\vec k,\beta ) & = & 3 g^2 \int d  \vec q
 F_{+}({\mid} \vec q\mid,\beta )\;
 F_{-}({\mid} \vec k +\vec q {\mid},\beta  ) \;
{\omega ({\mid  \vec  k +\vec q \mid},\beta )\over
\omega  ({\mid \vec  q\mid},\beta  )}
\nonumber \\
& \times & \bigl  (1+{{(q^2 +\vec k.\vec q)^2
}\over{q^2(\vec k+\vec q)^2}}\bigr )
\end{eqnarray}

\noindent  and
\begin{equation}
\phi  (k,\beta )=  {3g^2\over  {8 \pi  ^2}}
\int {{dk'}\over {\omega  (k',\beta  )}}\;F_{+}(k,\beta )
\biggl (  k^2+{k'}^2-{(k^2-{k'}^2)^2
\over{2kk'}}\log \Big | {{k+k'}\over{k-k'}}
\Big |  \biggr ).
\end{equation}
\noindent
The above equation
 is the parallel of equation (21) for zero temperature.
Further, $\omega(k,\beta )$ is as given in equation (34).

Using the form (22a) for $f(\vec k)$, we write the energy density,
$\epsilon_{0}(\beta)$ in terms of the dimensionless
quantities $x={\sqrt {B}}k$, $\mu={\sqrt B}m_{eff}(\beta)$
and $y={\beta\over {\sqrt B}}$ as
\begin{eqnarray}
\epsilon_{0}(A,\beta)
& = & {1\over {B^2}}(I_{1}(A,y)+I_{2}(A,y)+{I_{3}(A,y)}^{2}+
I_{4}(A,y)) \nonumber \\
& \equiv & {1\over {B^2}}F(A,y),
\end{eqnarray}
\noindent where
\begin{mathletters}
\begin{eqnarray}
I_{1}(A,y)
= {4\over {\pi^2}} & \int &  \omega(x)x^{2}dx
\Bigg[ \bigg(A^{2}e^{-x^2}
-Ae^{-{{x^2}\over 2}}(1+A^{2}e^{-x^2})^{1\over 2}\bigg)
\bigg(1+{2\over \exp{(y\omega(x,y))}-1}\bigg)\nonumber\\ & + &
{1\over \exp{(y\omega(x,y))}-1}\Bigg],
\end{eqnarray}
\begin{eqnarray}
I_{2}(A,y)=
{4\over {\pi^2}} & \int & {x^{4}\over \omega(x)}dx
\Bigg[ \bigg(A^{2}e^{-x^2}
+Ae^{-{{x^2}\over 2}}(1+A^{2}e^{-x^2})^{1\over 2}\bigg)
\bigg(1+{2\over \exp{(y\omega(x,y))}-1}\bigg)\nonumber \\ & + &
{1\over \exp{(y\omega(x,y))}-1}\Bigg],
\end{eqnarray}
\begin{eqnarray}
 I_{3}(A,y)=
{{2g}\over {\pi^2}} & \int &  {x^{2}\over \omega(x)}dx
\Bigg[ \biggl (A^{2}e^{-x^2}
+Ae^{-{{x^2}\over 2}}(1+A^{2}e^{-x^2})^{1\over 2}\bigg)
\bigg(1+{2\over \exp{(y\omega(x,y))}-1}\bigg)\nonumber\\
& + & {1\over \exp{(y\omega(x,y))}-1}\Bigg]
\end{eqnarray}
\noindent  and
\begin{equation}
I_{4}(A,y) =4\times {(2 \pi )^{-6}}\int d{\vec x}
{{G(\vec x ,y)}\over {x^2+\phi (x,y)}}
\end{equation}
\end{mathletters}
 with
\begin{eqnarray}
 G(\vec x,y)  =3g^2
& \int &  {d \vec x'}
\biggl [  \biggl (A^{2}e^{-{x'}^2}+Ae^{-{{{x'}^2}/2}}
{(1+A^{2}e^{-{x'}^2})}^{1\over 2}\biggr ) \nonumber \\
  & \times &
\biggl (1+{2\over \exp{(y\omega(x',y))}-1}\biggr )+
{1\over \exp{(y\omega(x',y))}-1} \biggr ]
\nonumber\\ & \times & \biggl [\biggl (A^{2}e^{-{(\vec x+\vec x')}^2}-
Ae^{-{{(\vec x +\vec x')}^2}/2}
{(1+A^{2}e^{-{(\vec  x+\vec x')}^2})}^{1\over 2}\biggr )\nonumber \\
& \times &
\biggl (1+{2\over \exp{(y\omega(\mid \vec x+\vec x' \mid,y)
)}-1}\biggr )+
{1\over \exp{(y\omega(\mid \vec x+\vec x' \mid,y))}-1} \biggr ]
\nonumber \\
& \times & {\omega (\mid \vec x +\vec x'\mid)\over {\omega ( x')}}
\biggl (1 +{{{({x'}^2 +\vec x .\vec x')}^2} \over
{\vec {x'}^2 (\vec x + \vec x')^2}}\biggr ),
\end{eqnarray}
 and
\begin{eqnarray}
\phi  (x,y) ={3g^2\over  {8 \pi  ^2}} & \int & dx'
 \biggl (  x^2+{x'}^2-{(x^2-{x'}^2)^2
\over{2xx'}}\log \Big | {{x+x'}\over{x-x'}}\Big | \biggr )
\nonumber \\ & \times & \Biggl [ \biggl (A^{2}e^{-{x'}^2}+Ae^{-{{{x'}^2}/2}}
{(1+A^{2}e^{-{x'}^2})}^{1\over 2}\biggr )
\biggl (1+{2\over \exp{(y\omega(x',y)
)}-1}\biggr ) \nonumber\\ & + &
{1\over \exp{(y\omega(x',y))}-1} \Biggr ].
\end{eqnarray}
\noindent In the above, now  $\mu$ becomes temperature
dependent i.e., $\mu=\mu(y)$. As before,
we determine the gluon mass, $\mu$ by identifying
the sum of the single contraction terms of the term
of ${\cal T}^{00}$ given by equation (8) which is
quartic in gauge field , as the mass term for the
gauge field. We thus have
\begin{eqnarray}
\mu(y)^2={{2g^2}\over {\pi^2}}
& \int &  {{x^2}dx\over \omega(x)}
\Biggl [ \biggl (A^{2}e^{-x^2}
+Ae^{-{{x^2}\over 2}}(1+A^{2}e^{-x^2})
^{1\over 2}\biggr )
\biggl (1+{2\over \exp({y\omega(x,y)})-1}\biggr ) \nonumber \\ & + &
{1\over \exp({y\omega(x,y)})-1}\Biggr ].
\end{eqnarray}

\noindent We note that $\mu (y)$ the output gluon mass in
dimensionless units, also occurs in $\omega(x,y)=(x^2+\mu (y)^2)^{1/2}$
inside the integration above for any fixed y. Thus it is a self
consistency equation which is the exact parallel of equation (27)
at zero temperature.

\section{Free energy extremisation and results}

In Ref.2, we had considered  extremisation of energy density
to obtain optimum $f(\vec k)$ describing vacuum structure.
However, at finite temperatures, the relevant quantity
for extremisation is free energy density, which
at any finite
temperature $T=1/\beta $ is given as \cite{tfd}

\begin{equation}
{\cal F}(A,\beta )=\epsilon _{0}-{1\over \beta }S.
\end{equation}
\noindent Here $\epsilon _{0}$ is as given in equation (45)
and $S$ is the entropy density given as \cite{tfd}
\begin{equation}
S=-2\times 8 \times (2\pi)^{-3}
\int d{\vec k}\Bigl(\sinh ^{2} \theta \; log \bigl(\sinh ^{2} \theta\bigr)
-\cosh ^{2} \theta \; log \bigl(\cosh ^{2} \theta\bigr) \Bigr )\end{equation}
\noindent The factor $2\times 8$ above comes from the transverse
and colour degrees of freedom for the gluon fields.
As before we may scale out the dimensional parameter ${1\over {B^2}}$
and write
\begin{eqnarray}
{\cal F}(A,\beta )& \equiv &  {1\over {B^2}}{F_1}(A,y)\nonumber
\\ & = &
{1\over {B^2}}\bigg [F(A,y)-{1\over y}{\cal S}(A,y)\bigg ],
\end{eqnarray}
\noindent where $F(A,y)$ is given in equation (45)
and ${\cal S}(A,y)$ is the entropy density
in dimensionless units given as
\begin{eqnarray}
{\cal S}(A,y)=-{8\over {\pi^2}} & \int &  x^{2}dx \biggl \{
\bigl ({1\over \exp{(y\omega(x,y))}-1}\bigr )
log \bigl ({1\over \exp{(y\omega(x,y))}-1}\bigr )\nonumber\\ & - &
\bigl (1+{1\over \exp{(y\omega(x,y))}-1}\bigr )
log \bigl (1+{1\over \exp{(y\omega(x,y))}-1}\bigr )\biggr \}.
\end{eqnarray}
\noindent We extremise $F_{1}(A,y)$  of equation (52) with
respect to A and obtain the optimum value of A as $A_{min}$ at
a given temperature T.
While extremising over A, the effective gluon mass $\mu (y)$ as
stated earlier is calculated in a self consistent manner using equation (49)
for any given $A$. $A_{min}$, so obtained is plotted as a
function of temperature in curve I of Fig.1 for the coupling
${g^2}/4\pi=0.8$. We find that $A_{min}$ decreases as we increase the
temperature and becomes zero at and above a critical value
$T_{c}$ around 275 MeV.
This value of $T_c$ may be compared with
the dissolution temperature \cite{karsh89}
$T_{dis}\simeq 210$ MeV above which
$c{\bar c}$ bound states dissociate. Such a
$J/{\psi}$ supression in QCD is
observed in the heavy ion collisions and is regarded as an
indication for the formation of quark gluon plasma.
We note that $T_c$ will change as the coupling constant changes,
and will be less for smaller coupling. The above value is
meant as an illustration to indicate features of strong coupling.

We next estimate the value of the SVZ parameter
using the value of $A_{min}$. This
at any finite temperature is given as
\begin{eqnarray}
& {{g^2}\over {4{\pi}^2}} & <:{G^a}_{\mu\nu}
G^{a\mu\nu}:>_{vac^{'},\beta }\nonumber \\
& {1\over {B^2}}{{g^2}\over {\pi^2}}
{\left(-I_{1}(A,y)+I_{2}(A,y)+I_{3}(A,y)^{2}-I_{4}(A,y)\right )\Big |
_{A=A_{min}}},
\end{eqnarray}
\noindent with $I_{1}(A,y)$, $I_{2}(A,y)$, $I_{3}(A,y)$
and $I_{4}(A,y)$ as in equations (46). Clearly this is the
parallel of equation (29) and gives the temperature dependance of
the SVZ parameter. The SVZ parameter which is plotted
 as curve II of Fig.1 is seen to
decrease with increase in temperature
and changes sign at temperature, $T=T_{c}$.
This may be compared with the recent hot sum rule calculations \cite{dom91}.
As in Ref.16, we have the SVZ parameter going to zero at a critical
temperature in contrast with the low temperature expansion
calculations \cite{gerb89}.
As earlier, in equation (54), the contributions from
$I_2$ and $I_3$  correspond to the magnetic sector, and,
$I_1$ and $I_4$, to the electric sector.
The magnetic contribution is always positive and varies slowly
with temperature. The contribution
from the electric sector varies more rapidly,
becomes negative and then overcomes that
from the magnetic sector leading to a change of sign of SVZ parameter.
The critical temperature, $T_c$ as may be
obtained from $A_{min}\rightarrow 0$ or SVZ parameter changing sign,
turns out to be the same and is around 275 MeV. Here these
 contributions are {\em not} associated with the masses of electric or
magnetic gluons. The present framework in fact does not consider formation
of electric or magnetic condensates separately. Such condensates occur
with a minimisation of energy density or free energy as earlier with
highly nontrivial contributions from equations (45) to (48) along with the self
consistency requirement of equation (49).

We also plot the effective gluon mass, $m_{eff}$ as a function
of temperature in curve III of Fig.1. Below the critical
temperature $T_c$, this parameter remains almost constant. However,
it increases linearly with temperature for
$T>T_{c}$, which corresponds to the calculations of the high
temperature limit \cite{anishet84}.
We may note that this gluon mass is a parameter
defined through $\omega (k,\beta )=\sqrt{k^{2}+m_{eff}(\beta )^{2}}$
in the thermal distribution function and is calculated self consistently
through equation (49).
Such a dynamically generated gluon mass has been
anticipated by Cornwall through Schwinger-Dyson equations in
QCD or through Monte Carlo calculations of gluon
propagators in lattice QCD \cite{cornwal82} at zero temperature.
 Similar effects are also seen at finite temperatures in lattice
 Monte Carlo calculations \cite{mandula88}.
 Although in all the calculations the word $``$gluon mass" is used,
the magnitude for the
same will depend on the  specific definition used in the calculation
and a difference in the corresponding numbers is to be expected.
A magnetic mass gap is called for as a
cure for the acute infrared divergence problem for finite
temperature perturbative QCD \cite{linde80}. This however gets
generated here dynamically in a natural way.

We plot the quantity ${\Delta \epsilon_{0} }/{T^4}\equiv ({\epsilon_{0} (T)
-\epsilon_{0} (T=0)})/{T^4}$ versus temperature in curve I of Fig. 2.
It increases continuously as T increases
 till $T_c$ and approaches a constant
value above $T_c$. This may be compared with the lattice calculations
at finite temperatures \cite{engels90,celik83}. Unlike these lattice results,
the small temperature
behaviour  here is not sharp. The reason may be because here
we have taken into account the nontrivial vacuum structure in QCD
through gluon condensates which is likely to be
important at small temperatures.
We note that the limit of $\Delta \epsilon /T^4$ in Fig.2
for $g^2/4\pi =0.8$ at
high temperature reached here is about 4.6 as compared to the
free gluons limit ${{\pi^2}\over{15}}(N^2-1)=5.26$ for N=3. We may also
remark our calculations not shown in the graph that for
$g^2/4\pi$=1, 0.8, 0.1, 0.01 the same expression become respectively
4.5, 4.6, 5.1 and 5.24 indicating that high temperarture is $not$ enough
to make the gluons behave like free particles \cite{engels90}.
We may also remark that
lattice calculations the above ratio falls short of 5.26 by a similar
margin \cite{celik83}.

 We next  plot $S/T^{3}$ from equation (51)
as a function of temperature
in curve II of Fig. 2,
which is seen to have similar behaviour as $\Delta \epsilon_{0} /T^{4}$
versus temperature except that the rate of rise is sharper than the
earlier curve for temperatures below $T_c$.

We now calculate the pressure P given as \cite{fetter}
\begin{equation}
P(\beta )=-{\cal F}(A,\beta )\Big |_{A=A_{min}}.
\end{equation}
To compare with the lattice calculations \cite{engels90}, we plot
the quantity
$(\Delta \epsilon_{0} -3 \Delta P)/{T^4}
\equiv {\big [(\epsilon_{0} -3P)(T)-(\epsilon _{0}-3P)(T=0)\big ]}/{T^4}$
as a function of temperature in curve III in Figure 2.
We may note that as in Ref.19 the deviation of the above quantity
from ideal gas behaviour $(\epsilon -3p)/T^4=0$ is pronounced
even for temperatures above $T_c$. As stated earlier, this probably
indicate that at high temperartures the effects of interactions
does not disappear, and nonperturbative methods like ours
or in lattice QCD, which always includes interactions, reflects this.
However, below $T_c$ the behaviour is different from that
of Ref.19. This may be due to the nontrivial vacuum structure
considered in the present framework, or a limitation of our
ansatz.

With a critical temperature as determined here, it is important to
recognise whether the phase transition is first order or second order.
As shown in the figures it appears to be a second order phase
transition with zero latent heat. However, since the calculations
were numerical some remarks regarding the same seem to be pertinent.
Near critical temperature the sensitivity of the calculations
entering through $A_{min}$ plotted as curve I of Fig.1. progressively
diminishes. They could be consistent with a first order phase transition
with a very small latent heat which our numerical calculations was
not able to resolve.

\section{ Discussions}

We have discussed here ground state in QCD in Coulomb gauge
at finite temperature through
a variational method.
Here we make no assumption regarding the temperature being
either high \cite{anishet84} or low \cite{gerb89}.
At each temperature, calculations
are done in a selfconsistent manner with the extremisation of
the free energy. These calculations are seen to go over to the zero
temperature limit \cite{am91} smoothly.
We note that an unusual feature here is the
generation of a masslike parameter for the gluons, which however
has been noted earlier in QCD for continuum \cite{cornwal82}
as well as lattice calculations \cite{mandula88}.

It is important to note that we have chosen here Coulomb gauge,
so that the conclusions are $not$ gauge invariant. The
dynamics of the vacuum structure in QCD is highly nontrivial.
We have concentrated on developing a nonperturbative
method for this. Here with Coulomb gauge only the transverse
gluons are dynamical. This has allowed us to obtain an
approximate solution for the vacuum structure. However, it will be
$necessary$ to choose different gauges and see which conclusions
remain unaltered.

An important feature of QCD of low temperature phase is confinement
as seen through area law behaviour of Wilson loops. Such a line
integral through Stokes theorem in simple cases may be converted to an
integral over $<{G^a}_{\mu\nu}{G^a}^{\mu\nu}>$ \cite{comm}, and the
nonvanishing of the same at low temperature phase
shall imply confinement.
An explicit calculation with the present model to determine
 the string tension or confinement \cite{nadkarni88,polya77} however
seems to be nontrivial  and has not been attempted
here.

A basic limitation of the present approach is the absence of a
running coupling constant. As discussed earlier, for such an approach
we need to calculate higher order amplitudes which we have not done
here. Getting a finite masslike term here leads us to suspect that
the infinities of renormalisation theory could be an artifact of perturbative
approach but such a statement should be associated with a more
serious calculations not attempted here. We have thought that
considering running coupling constant can only go with such calculations.
We further note that we would have liked to associate the scale parameter
of vacuum structure with $\Lambda_{QCD}$, instead of doing so with
SVZ parameter. The same could not be achieved here since $\Lambda_{QCD}$
gets determined through renormalisation group equation.

An important aspect of the present
methodology is that phase transition is treated here quantum mechanically
as a realignment of QCD vacuum, instead of being obtained
by the extremisation in a classical effective theory.
The equations for QCD are solved at the level of quantum
mechanical expectation values with respect to the ground state,
and $does$ $not$ correspond to taking
classical mean fields. The method is nonperturbative
and variational. Although it is limited by the ansatz of the ground
state, the strong coupling features as in lattice gauge
theory calculations show up. Thermofield method enables us to
replace statistical averages by expectation values in an enlarged
Hilbert space. This feature has been exploited by us to study the
nontrivial vacuum structure at zero temperature as well as QCD
ground state at finite temperature in one framework.

\acknowledgements

The authors are thankful to J. Pasupathy, J.C. Parikh, N. Barik,
Snigdha
Mishra, and P.K. Panda for many useful
discussions. One of the authors (AM) would like to thank
the Council of Scientific and Industrial Research (C.S.I.R)
for a fellowship. SPM would like to thank Department of
Science and Technology, Government of India for
research grant no SP/S2/K-45/89 for financial assistance.

\figure
{ We plot here $A_{min}$,
$SVZ\equiv {{g^2}\over {4\pi^{2}}}<:{G^a}_{\mu \nu }G^{a \mu \nu }:>
_{vac';\beta }$
 and gluon mass $m_{eff}$ respectively in curves I, II and III
 as functions of temperature, T in MeV.
In curve II, SVZ is expressed in units of $10^{-2}GeV^{4}$,
 and, in curve III, effective gluon mass $m_{eff}$ is expressed
in units of 200 MeV. Each of these curves leads to a critical
temperature, $T_{c}\simeq 275MeV$.\label{fig1}}
\figure{ We plot here
$\Delta \epsilon _{0}/T^{4}$, $S/T^{3}$
and $(\Delta \epsilon _{0}-3 \Delta P) /T^{4}$
in curves I, II and III respectively as a
function of temperature, T in MeV.\label{fig2}}
\end{document}